# Radar Inband and Out-of-Band Interference into LTE Macro and Small Cell Uplinks in the 3.5 GHz Band

Mo Ghorbanzadeh, Eugene Visotsky, Prakash Moorut, Weidong Yang
Nokia Solutions and Networks US LLC
Arlington Heights, USA
{mo.ghorbanzadeh, eugene.visotsky, prakash.moorut, weidong.yang}@nsn.com

Charles Clancy
Hume Center for National Security and Technology
Virginia Tech, Arlington, USA
tcc@vt.edu

*Abstract*—National Telecommunications and Information Administration (NTIA) has proposed vast exclusions zones between radar and Worldwide Interoperability for Microwave Access (WiMAX) (WiMAX) systems which are also being considered as geographic separations between radars and 3.5 GHz Long Term Evolution (LTE) systems without investigating any changes induced by the distinct nature of LTE as opposed to WiMAX. This paper performs a detailed system-level analysis of the interference effects from shipborne radar systems into LTE systems. Even though the results reveal impacts of radar interference on LTE systems performance, they provide clear indications of conspicuously narrower exclusion zones for LTE vis-à-vis those for WiMAX and pave the way toward deploying LTE at 3.5 GHz within the coastline populous areas.

## I. INTRODUCTION

Spectrum sharing is an elegant solution to the dramatic increase in the data traffic volume of mobile broadband networks over the next 20 years [1]. It helps meet demands of Mobile Network Operators (MNOs) by assigning them new pieces of spectrum. A pioneering effort toward realizing the spectrum sharing was made by the Council of Advisers on Science and Technology (PCAST) [2] to leverage the full potential of the government-held spectrum. PCAST spurred the Federal Communications Commission (FCC) to issue a Notice of Proposed Rulemaking (NPRM) [3] to designate the 3550 - 3650 MHz range, abbreviated as the 3.5 GHz band, for mobile broadband. In sequel, the National Telecommunications and Information Administration (NTIA) recognized radar systems as major band incumbents and conducted a measurement campaign [4], which revealed a low average temporal utilization of the band by the incumbents and the potential for spectrum sharing.

However, an effective spectrum sharing refrains from destructive interference between incumbents and entrants. Peculiarly to the 3.5 GHz band, NTIA's investigated the interference between radar and Worldwide Interoperability for Microwave Access (WiMAX) systems [5] and suggested exclusion zones reaching 557 km inland (Figure 1) [6] where no 3.5 GHz communications systems, e.g. Long Term Evolution (LTE) [7] system can be deployed. This hinders deploying the 3.5 GHz LTE in coastal regions of the United States (U.S.) where over 55% of the American reside [8].

Henceforth, judiciously reducing the exclusion zones affords MNOs to utilize this band to enhance mobile broadband coverage. The geographic separations in [6] simplistically emerged from link budget analyses of radar-WiMAX ecosystems, while LTE is the expected cellular technology in the 3.5 GHz band and considering the nuances of the LTE link-level protocol (turbo coding, advanced scheduling, Hybrid Automatic Repeat Request - HARQ, etc.) can alter the exclusion zones. So, analyzing radar-to-LTE interference (not WiMAX) creates relevant exclusion zones in the 3.5 GHz band. Besides, small cell implementation of LTE is becoming more popular as it escalates the network capacity and extends the service coverage.

This paper looks into the interference from S-band rotating radars into the uplink of a Time Division Duplex (TDD) LTE. The investigation relies on macro and small cell LTE system level simulations compliant to the 3rd Generation Partnership Project (3GPP) [9], simulating radars with operating parameters from NTIA [6], free space path loss (FSPL) [10], and irregular terrain model (ITM) diffraction loss [11]. The simulations show that LTE macro and small cells are vulnerable to cochannel and out-of-band radar interference, but they reduce the exclusion zones significantly less than the NTIA report [6]. Furthermore, the impact from radar into the small cells in the uplink direction is more pronounced than the macro cells. Furthermore, the out-of-band interference into LTE macro and small cells is not negligible. To our best knowledge, no prior works on radar-to-LTE interference render a full consideration to radar on-off and rotation interference characteristics along with LTE deployments protocol details.

### A. Related Work

Realizing that the cellular system-radar spectrum sharing necessitates mitigating their mutual interference, work in [4] investigated the WiMAX-radar mutual interference and concluded that large geographic separations between the two systems are required, precluding WiMAX deployability in the coastline. Cotton et. al. [10] performed tests, using an S-band shipborne radar in San Diego littoral areas, measuring fortnight's temporal band occupancy and found that the 3.5 GHz spectrum is not often occupied by radar transmissions, thereby underlining the promising potential of the germane band for spectrum sharing. Lackpour et. al. [11] suggested a general spectrum sharing scheme based on time, space, frequency, and system-level modifications, of which the last is inconducive to real-world implementation. Khawar et. al. [12] proposed projecting a radar signal onto the null space of the interference channel to allow for spectrum sharing with futuristic radars. Sanders et. al. [13] performed an experiment with cabled RF connections to observe the effects of interference from radar waveforms onto a proprietary 3.5 GHz LTE evolved NodeB (eNB). In particular they considered the throughput loss and block error rate (BLER) for the uplink LTE system; however, their results were varied as some waveforms did not have any appreciable effect on the LTE while others undermined the performance drastically. They did not consider any propagation models, nor did they perform any simulation of a more realistic radar or LTE system.

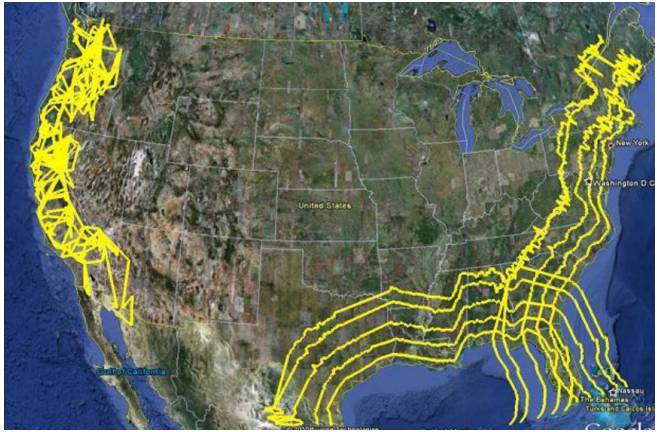

**Figure 1: Radar-WiMAX exclusion zones, spanned by the yellow curves, exceed 500 km [6].**

*B. Organization*

The rest of this manuscript proceeds with presenting our simulation setup in section II and simulation results in section III. Section IV concludes the paper.

## II. EXPERIMENTS

We set a simulation time 5 seconds to radiate radar signals onto the Base Stations (BSs) of an LTE macro and small cell system. First, the radar is cochannel with the LTE network; then, it offsets 5 and 10 MHz from the LTE channel. The radar is positioned 50, 100, 150, and 200 km away from the LTE systems and the interference is simulated. Next, section II.A presents the simulations germane to the radar system.

*A. Radar Simulation*

The radar parameters are listed in Table 1 based on NTIA [6]. The items marked with asterisk were not disclosed in [6] due to their tactical sensitivity and were set typical operating values for medium-to-large shipborne S-band radars [12]. Being 50, 100, 150, and 200 km away from the LTE, radar emanates pulses in to the communications network as in Figure 2 (a). Scanning 360 deg azimuthally with 30 rotations per minutes (rpm), the radar single rotation time becomes 2 s during which 4000 pulses each 83 dBm are emanated as the pulse repetition interval (PRI) is 0.5 ms (1/0.5 ms = 2000 Hz). The horizontal beamwidth 0.81 deg necessitates 445 beam positions to cover the 360 deg search fence [12]; so the antenna dwell time grows 4.5 ms in which 9 pulses as in Figure 2(b) are transmitted by the radar. Here, the abscissa and ordinate axes represent time in seconds and amplitude in Volt respectively.

The radar antenna has cosine pattern with equation (1) [6] plotted in Figure 3 as the normalized gain in terms of the off-boresight angle θ. Here, the first, second, and third expressions are the theoretical directivity pattern, a mask equation for pattern deviation from the theoretical value one at the side-lobe (-14.4 dB main beam), and back-lobe, respectively. As we can see, the back-lobe is fixed at 50 dBm below the main lobe.

$$G(\theta) = \begin{cases} \dfrac{\pi}{2} \left( \dfrac{\cos(\dfrac{68.8\pi \sin(\theta)}{\theta_{3dB}})}{(\dfrac{\pi}{2})^2 - (\dfrac{68.8\pi \sin(\theta)}{\theta_{3dB}})^2} \right) \\ -17.51\log_e \left( \dfrac{2.33|\theta|}{\theta_{3dB}} \right) \\ -50\,dB \end{cases} \quad (1)$$

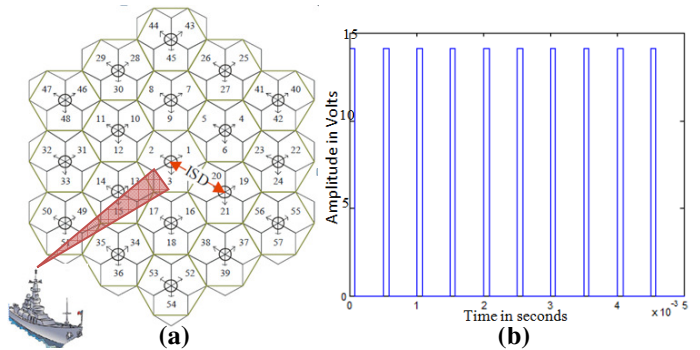

(a)      (b)

**Figure 2: (a) Simulation scenario subsumes a radar next to the 3.5 GHz LTE. (b) Radar pulses radiated on the LTE during the antenna dwell time.**

**Table 1: Radar parameters from [6], except those marked with * where we used typical parameters.**

| Parameters | Value |
|---|---|
| Operating Frequency | 3.5 GHz* |
| Peak Power | 83 dBm |
| Antenna Gain | 45 dBi |
| Antenna Pattern | Cosine |
| Antenna Height | 50 m |
| Insertion Loss | 2 dB |
| Pulse Repetition Interval | 0.5 ms |
| Pulse-Width | 78 µs |
| Rotation Speed | 30 rpm* |
| Azimuth Beam-Width | 0.81 deg* |
| Elevation Beam-Width | 0.81 deg* |
| Azimuth Scan | 360 deg |
| Pulse Repetition Interval | 0.5 ms |
| Distance to LTE | 50, 100, 150, 200 km |

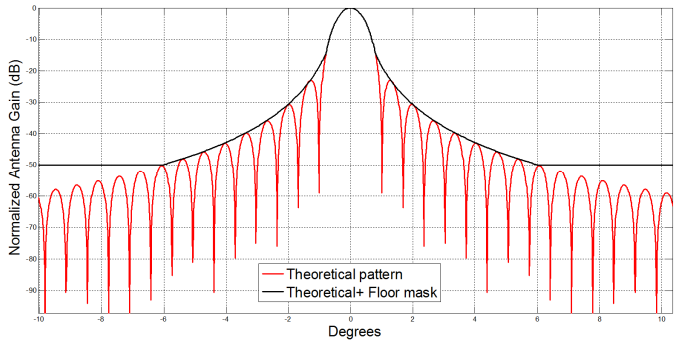

**Figure 3: Radar Antenna Pattern [6]:** The black curve represents the pattern after the first side lobe is 14.4 dB below the main lobe. The back lobe is constant at -50 dB.

*B. LTE Simulation*

The LTE system level simulation is 3GPP compliant [9,13] and based on the International Telecommunications Union (ITU) recommendations on International Mobile Telecommunications-Advanced (IMT-A) radio interface technologies [14]. It contains indoor, small cell and macro cell infrastructures and has a full-buffer traffic model. Users are simulated at a pedestrian velocity 3 km/h. First, we used a macro cell model - urban macro (UMa) - with an intersite distance 500 m. We deployed 7 sites with 120 deg sectors, and the BS antenna pattern per sector is given in equation (2) [14] where $G_A$ and $\theta_A$ ($G_E$ and $\theta_E$) are the antenna azimuth (elevation) pattern and azimuth (elevation) angle off the boresight. Here, -180° ≤ $\theta_A$ ≤ 180° (-90° ≤ $\theta_E$ ≤ 90°), antenna azimuth (elevation) downtilt is $\theta_{A,t} = 0°$ ($\theta_{E,t} = 15°$), $A_m = 20$ dB is the maximum attenuation, and $\theta_{3dB}$ is the antenna 3dB beamwidth. The composite antenna pattern is expressed in equation (3), plotted in Figure 4.

$$G_i(\theta_i) = -\min\{12(\frac{\theta_i - \theta_{i,t}}{\theta_{3dB}})^2, A_m\}, i \in \{A, E\} \quad (2)$$

$$G = -\min\{-(G_A(\theta_A) + G_E(\theta_E)), A_m\} \quad (3)$$

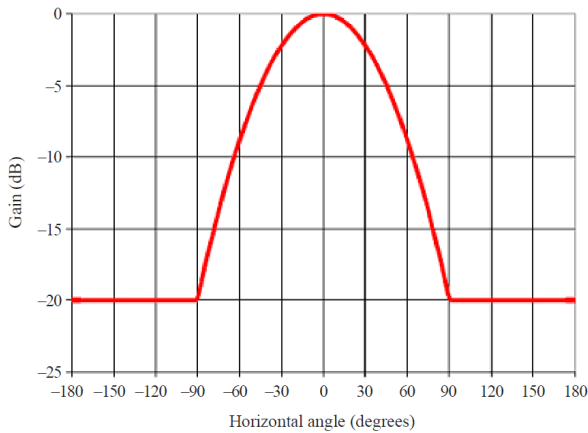

**Figure 4:** Composite antenna pattern the eNBs for 3-sector cells from equation (4).

In addition, our LTE simulation included small cells – urban micro (UMi) – whose main distinction from the macro cells lies in the fact that the cells are not sectorized, they use omnidirectional BS antennas (therefore of 0 dB gain), and their antenna heights are 10 m as opposed to the 25 m in the UMa. The LTE parameters for both macro cell and small cell LTE simulations are listed in Table 2.

**Table 2: LTE Simulation Parameters [9, 14].**

| Parameters | Value |
|---|---|
| Operating Frequency | 3.5 GHz |
| Layout | Hexagonal grid |
| Mode | TDD (In each TDD cycle, the uplink traffic ran for 3 ms of on-time, with a 2 ms off-time interval for downlink traffic (DL:UL ratio 2:3) |
| Macro/Small Cells BS TX Power | 46/30dBm |
| UE Transmit (TX) Power | 23 dBm |
| Macro-cell Sites, Cells | 7, 21 |
| Small Cells | 84 (4 per macro cell area) |
| Indoor UE for Macro, Small cells | 80%, 20% |
| Bandwidth for Macro, Small cells | 20 MHz |
| Macro, Small Cells BS Antenna Gain | 17, 5 dBi |
| UE Antenna Gain | 0 dBi |
| Intersite Distance for Macro | 500 m |
| Minimum UE-BS Distance for Macro, Small Cells | 25, 5 m |
| BS Antenna Downtilt | 12 deg |
| BS/UE Antenna Height | 25/1.5 m |
| UE Distribution | Uniform |
| UE Mobility | 3 km/h, uniform direction |
| BS/UE Noise Figure (NF) | 5/9 dB |
| Thermal Noise | -174 dBm/Hz |
| Service Profile | Full buffer best effort |
| UE per for Macro, Small Cell | 10, 30 |
| Channel Model for Macro, Small Cells | UMa / UMi |

*C. Radar to LTE Propagaion Model*

The radar signals undergoes a propagation loss before they reach the LTE BSs, and appropriate models relying on the distance and terrain between the radar and LTE should be leveraged to obtain how strong the radar pulses are once they arrive at the BSs. For the line-of-sight (LoS) and non-LoS (NLoS) regions, we respectively use the FSPL and ITM, predominantly used models by the FCC and NTIA. FSPL is expressed in equation (4) [10] where $f$ is the radar operating frequency in hertz, $r$ is the distance in km at which FSPL $L_{dB,FSPL}$ in dB is given, and $r_{LoS}$ is the LoS region border in km as equation (5) [2] where $h_{radar}$ and $h_{LTE}$ is the radar and LTE antenna height as 50 m and 25 m for macrocells (10 m for small cells).

$$L_{dB,FSPL}(r) = 20\log(f) + 20\log(r) + 32.45, r < r_{LoS} \quad (4)$$

$$r_{LoS} = 4.1(\sqrt{h_{radar}} + \sqrt{h_{LTE}}) \quad (5)$$

As for the diffraction loss in the NLoS region, we leverage the ITM in its area prediction mode (APM) [11] with the terrain roughness 10 and 20 m, LTE macro and small cell antenna heights 25 and 10m, radar antenna height 50 m, ground dielectric constant 15, ground conductivity 0.005 S/m, refractivity 301 N-units, continental temperate climate, and single message mode as in Table 3. The plots for the FSPL and ITM diffraction losses are depicted in Figure 5, where the green curve indicates FSPL, red curve is the diffraction loss for the macro cell scenario, and blue curve is the diffraction loss for the small cell scenario as a function of the travelled distance by radar pulses. It is noteworthy that the two models predict very close values for the losses in the LoS region, approximately 50 km, whereas this loss sharply elevates in the NLoS region, where ITM model proves valid.

**Table 3: ITM Parameters (adopted from [4]).**

| Parameters | Value |
|---|---|
| Operation Mode | Area Prediction Mode (APM) |
| Macro, Small Cells LTE/Radar Antenna Height | 10, 25/50 m |
| Dielectric Constant | 15 |
| Conductivity | 0.005 S/m |
| Refractivity | 301 N-units |
| Climate | Continental Temperate |
| Variability Mode | Single Message |

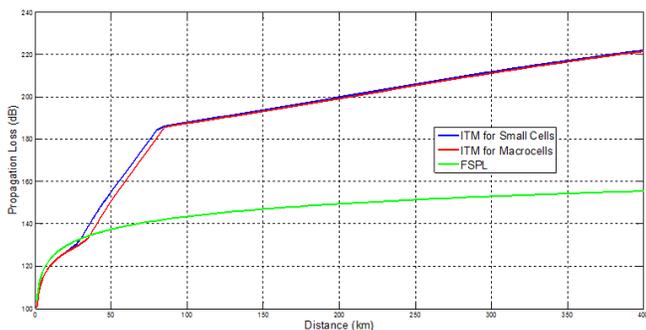

**Figure 5: LoS FSPL in green and NLoS ITM loss in blue and red for macro and small cells respectively represent the extent the radar signal degrades vs. the traveled distance.**

### III. SIMULATION RESULTS

We set the simulation time to 5 seconds, during which the the impact of the radiation of the radar with parameters in section II-A onto the eNBs of the LTE cellular system with parameters in section II-B, is investigated. The radar is cochannel with the LTE system and rotates 360 deg in azimuth in 445 beam positions where the radar sojourns for the dwell time 4.5 ms and sends 9 pulses 78 µs wide and 83 dBm through its 45 dBi antenna to the eNBs in the beam position based on equation (1). For instance, at the LTE system, the width of the radiation spans 1.5, 3.0, 4.5, and 6.0 km when the radar is respectively 50, 100, 150, and 200 km away from the LTE system. As such, all the eNBs covered by the radiation width would suffer from the radar pulses which are amplified by the transmitter and antenna and undermined by the propagation losses explained in section II-c.

#### A. Radar to LTE Macrocell BSs Interference

First, we look at the macrocell results. As we can see from Figure 1, slight throughput losses incurred when the radar is 50, 100, 150, and 200 km away vis-à-vis the baseline. It is noteworthy that Figure 1 represents relative values.

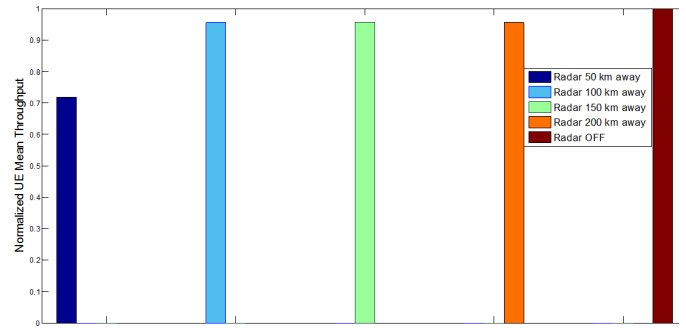

We plot the signal-to-interference-to-noise ratio (SINR) of an LTE macro BS versus LTE symbol and subcarrier indices in Figure 2 where we observe an SINR drop due to the radar pulse affecting LTE symbols on the uplink during the simulation time. Interestingly, even when the radar is present, the SINR recovers back to its normal baseline situation until the next pulse hits the same region (same beam position). Because the radar pulse is assumed to be centered in the LTE band, most of the pulse energy is concentrated around subcarrier 300 (in the middle of the LTE band). Also, at 78 □s, the pulse slightly exceeds the duration of the LTE symbol (71.4 □s). Thus, most of the energy is concentrated in symbol 1 and symbol 8, with some remaining pulse energy also present in symbols 2, 9 and 14. This is promising as only certain symbols during an LTE sub-frame are affected by the radar signal.

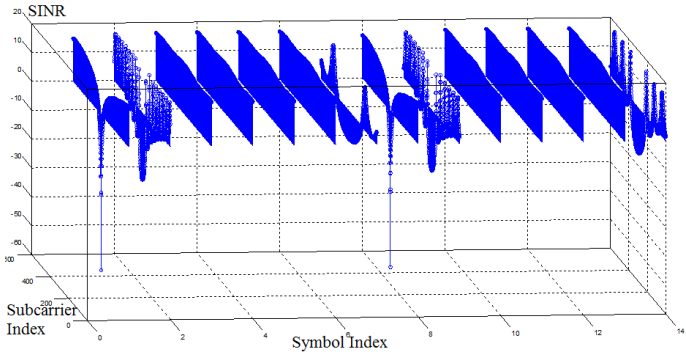
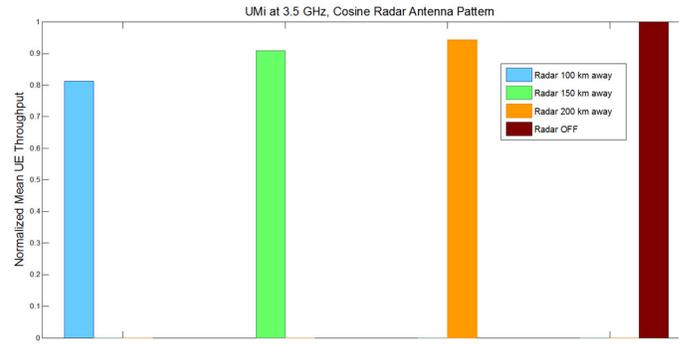

Moreover, Figure 7 does not precisely show the slight throughput losses incurred when the radar is 100, 150, and 200 km away vis-à-vis the baseline. In order to reveal the throughput degradation for the aforesaid distances, we resort to the cumulative distribution functions (CDF)s of the germane throughput losses, depicted in Figure 8. Here, the blue curve represents the baseline, i.e. the situation where no radar is present in the vicinity of the LTE system. However, as we can observe from the cyan, green, and black curves, the radar at distances of correspondingly 200, 150, and 100 km away from the LTE system generates some throughput loss. On ther other hand, the throughput degradation is more pronounced when radar is 50 km away from the LTE system (red curve). It is to be noted that the abscissa values are omitted due to confidential considerations from Nokia. However, as an illustration, for the values intersected by the dashed vertical line, when the radar is 50 km away, more than 50% of the UEs have a throughput less than the corresponding values obtained by the intersection of the abscissa and the dashed line. In contrast, less than 30% of the UEs degrade to the same throughput when the radar is 100, 150, and 200 km away from the shore.

*B. Radar to LTE Small Cells BSs Interference*

Next, we explain the results for the small cells. As we can observe from the blue, green and orange bars in Figure 3, when radar is 100, 150, and 200 km away from the LTE system, there is a slight UE throughput decline. As it is expected, the further away the radar, the higher the mean UE throughput as interference becomes less pronounced due to the diffraction loss caused by the ITM in the NLoS region. It is noteworthy that Figure 3 represents relative values. Therefore, for small cells and for radars 100 km away from the cellular system, the impact on LTE is modest. The brown bar represents the baseline, i.e. no radar operates in the vicinity of the cellular system. The investigation for radar distances less than 100 km away from the small cells is currently undergoing.

Finally, we plot the SINR of an LTE BS versus LTE symbol for small cells and subcarrier indices in Figure 4 where we observe an SINR drop due to the radar pulse affecting LTE symbols on the uplink during the simulation time. Interestingly, even when the radar is present, the SINR recovers back to its normal baseline situation until the next radiation hits the same region (same beam position). Being 78 µs wide, the pulse exceeds the duration of the LTE symbols (71.4 µs).

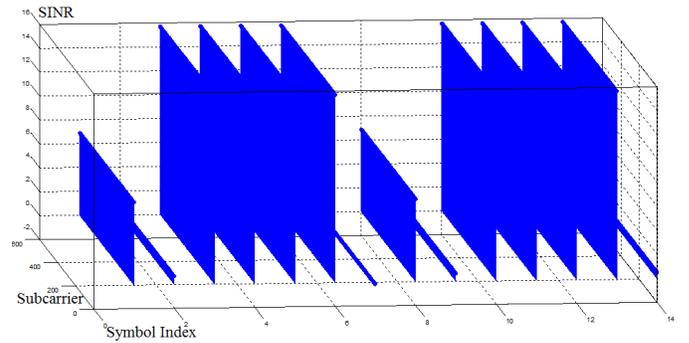

IV. CONCLUSION

In this paper, we studied the impact of shipborne S-band radar systems that are co-channel with and in the vicinity of a cellular 3.5 GHz LTE systems. We leveraged the LTE standard to simulate LTE uplink at a system level. Furthermore, we deployed radar parameters according to the NTIA report [4] as a bench mark in order to simulate similar radar systems.

Moreover, we simulated free-space and ITM pathlosses conducive to modeling the LoS and diffraction losses attenuating the radiated signals from the radar system to obtain the relevant signal levels at the LTE system. In the simulations, we assessed the radar impact by observing the SINR for symbol and subcarrier indices. We observed that the presence of the radar reflects SINR plummets for the eNBs during the antenna dwell time where an eNB is radiated. Furthermore, we looked at the UE throughputs losses when a radar interference occurs. Contrasting the baseline with interference scenarios at various distances between the radar and LTE system, we realized that even though the radar interference is clearly visible by degrading the UE throughputs in the uplink, the throughput loss is tolerable even with radar deployed only 50 km away from the LTE system.

Henceforth, in view of the results of the current article, the authors propose a significant reduction in the aforementioned exclusion zones. Such a measure can constructively afford cellular providers to share the 3.5 GHz band with government and expand their services to the coastline metropolitan areas, which not only ensues conspicuous revenues for the providers, but also motivates fulfilling spectrum sharing with government in the other bands as well.